\documentclass[english]{IEEEtran}
\usepackage[T1]{fontenc}
\usepackage[latin9]{inputenc}
\usepackage{bm}
\usepackage{amsthm}
\usepackage{amsmath}
\usepackage{amssymb}
\usepackage{graphicx}

\makeatletter
\theoremstyle{plain}
\newtheorem{thm}{\protect\theoremname}
\theoremstyle{remark}
\newtheorem{rem}[thm]{\protect\remarkname}

\makeatother

\usepackage{babel}
\providecommand{\remarkname}{Remark}
\providecommand{\theoremname}{Theorem}

\begin{document}

\title{Power Allocation in Compressed Sensing of Non-uniformly Sparse Signals}

\author{Xiaochen Zhao and Wei Dai\\
Department of Electrical and Electronic Engineering\\
Imperial College London, United Kingdom}
\maketitle
\begin{abstract}
This paper studies the problem of power allocation in compressed sensing
when different components in the unknown sparse signal have different
probability to be non-zero. Given the prior information of the non-uniform
sparsity and the total power budget, we are interested in how to optimally
allocate the power across the columns of a Gaussian random measurement
matrix so that the mean squared reconstruction error is minimized.
Based on the state evolution technique originated from the work by
Donoho, Maleki, and Montanari, we revise the so called approximate
message passing (AMP) algorithm for the reconstruction and quantify
the MSE performance in the asymptotic regime. Then the closed form
of the optimal power allocation is obtained. The results show that
in the presence of measurement noise, uniform power allocation, which
results in the commonly used Gaussian random matrix with i.i.d. entries,
is not optimal for non-uniformly sparse signals. Empirical results
are presented to demonstrate the performance gain.
\end{abstract}

\section{Introduction}

Compressed Sensing has been widely studied to reconstruct sparse signals
from underdetermined observations \cite{Donoho06compressedsensing}.
The observation $\bm{y}\in\mathbb{R}^{m}$ is measured from the linear
model 
\begin{equation}
\bm{y}=\bm{A}\bm{x}+\bm{w},\label{eq:CS}
\end{equation}
 where $\bm{A}\in\mathbb{R}^{m\times n}$ $\left(m<n\right)$ is the
measurement matrix, $\bm{x}\in\mathbb{R}^{n}$ is the unknown sparse
signal, and $\bm{w}\in\mathbb{R}^{m}$ is the white Gaussian noise
with covariance $\sigma^{2}\bm{I}$. In this paper, we are particularly
interested in non-uniformly sparse signals where different signal
components may have different nonzero probabilities. Such signals
arise in many practical scenarios. For example, in the multiple-source
localization problem, the sources (corresponding to nonzero signal
components) are often clustered in certain areas. For natural images,
the nonzero wavelet coefficients form a tree structure \cite{5437428}.
In video surveillance, the signals from adjacent frames share many
nonzero components \cite{5205717}. Using the non-uniformly sparsity
appropriately can help improve the compressed sensing reconstruction
performance, see \cite{conf/isit/OymakKH12,conf/isit/Kha09,5712950,6409458}
for examples. 

In this paper, we focus on the measurement matrix design problem when
non-uniformly sparse signals are involved. More specifically, given
a total power budget, we are interested in the optimal power allocation
across the columns of a Gaussian random measurement matrix to minimize
the reconstruction error. Similar problems have been considered in
the adaptive sensing setup where non-uniformly sparse statistics are
generated in the initial sensing process and that information is used
to design the measurement matrices in later stages. Examples include
\cite{haupt2009adaptive}, \cite{journals/jmlr/Seeger08}, \cite{ErvinTanczos},
and \cite{2013ISTSP...7..783W}, to name a few. Different from adaptive
sensing, we assume that the non-uniformly sparse statistics are given
a priori, which can be viewed as a simplification of adaptive sensing.
As we shall show later, this simplification allows a closed form formula
to compute the asymptotically optimal power allocation policy under
certain assumptions. 

Our technique originates from the so-called approximate message passing
(AMP) algorithm and the associated analysis developed by Donoho et
al. \cite{domamo09}. AMP assumes no power allocation, that is, the
entries of the measurement matrix are generated from i.i.d. Gaussian
random variables. The key element of the theoretical analysis is the
so called state evolution. It quantifies exactly the under-sampling
rates when perfect reconstruction is possible (referred as the phase
transition curve \cite{Donoho06compressedsensing}), or the worst-case
reconstruction mean squared error (MSE) for a given noise variance
(referred to as minimax MSE) \cite{journals/tit/DonohoMM11}. The
same technique has been applied to non-uniformly sparse signals in
\cite{5712950} and block separable signals in \cite{6409458}, and
also been extended to more general channel models \cite{GAMP,Fixed_GAMP}.
With power allocation, the measurement matrix in this paper does not
contain i.i.d. Gaussian entries. It can be viewed as special cases
of the generalised channel model. 

The main contribution of this paper is the asymptotically optimal
power allocation to minimize the reconstruction MSE. More specifically,
we revise the standard AMP algorithm to accommodate non-uniformly
sparse signals and Gaussian measurement matrices with power allocation.
The reconstruction MSE of the revised AMP algorithm has been exactly
quantified in an asymptotic regime. Based on it, the asymptotically
optimal power allocation policy is derived. Note that the presented
analysis is mainly for the worst case as it results in closed-form
formulas. The analysis can be generalised for more practical scenarios
with minor modifications and produce satisfactory results according
to our simulations.

\section{\label{sec:2} Problem Formulation and Preliminaries}

In standard compressed sensing (CS) settings, the entries of the measurement
matrix $\bm{A}$ are generated from i.i.d. Gaussian random variables.
However, this may not be optimal in terms of reconstruction distortion
when the unknown signal $\bm{x}$ is non-uniformly sparse, i.e., the
probabilities for different entries to be nonzero may be different.
Consider the example where $\bm{x}=[\bm{x}_{{\cal I}_{1}},\;\bm{x}_{{\cal I}_{2}}]$
and the entries in $\bm{x}_{{\cal I}_{1}},\;\bm{x}_{{\cal I}_{2}}\in\mathbb{R}^{n/2}$
have different nonzero probabilities. In an extreme case, suppose
that the entries in $\bm{x}_{{\cal I}_{1}}$ share the same prior
distribution with strictly positive nonzero probability while all
the entries in $\bm{x}_{{\cal I}_{2}}$ are zeros. Fix the total power
budget, i.e., the squared $\ell_{2}$-norm of each row of the measurement
matrix is fixed to a constant. Different from the equal power allocation
in standard CS, a more sensible way is to spend no sensing power on
the zero components in $\bm{x}_{\mathcal{I}_{2}}$ but allocate all
sensing power evenly to the columns corresponding to $\bm{x}_{\mathcal{I}_{1}}$. 

The formal setting is as follows. Let 
\begin{equation}
{\cal F}_{\epsilon}=\left\{ p:\; p\left\{ 0\right\} =1-\epsilon\right\} \label{eq:prob-family-epsilon}
\end{equation}
be the family of probability distribution with a mass $1-\epsilon$
at zero. Assume a block-sparsity signal $\bm{x}=\left[\bm{x}_{{\cal I}_{1}};\,\bm{x}_{{\cal I}_{2}};\,...;\,\bm{x}_{{\cal I}_{s}}\right]$
where $p_{\epsilon_{i}}\in\mathcal{F}_{\epsilon_{i}}$ and $p_{\epsilon_{i}}=p_{\epsilon_{j}}$
if $i,\; j\in\mathcal{I}_{k}$, $k\in\left[s\right]$. For the purpose
of power allocation, suppose that each column of $\bm{A}$, denoted
by $\bm{A}_{i}$, $i\in\left[n\right]$, contains entries generated
from i.i.d. Gaussian random variables with ${\cal N}\left(0,\;\sigma_{i}^{2}/m\right)$.
Fix a total power budget $\Sigma_{i=1}^{n}\sigma_{i}^{2}=n$. The
goal is to minimize the reconstruction error subject to the total
power budget,

\begin{equation}
\underset{\sigma_{1}^{2},\cdots,\sigma_{n}^{2}}{\min}\;\frac{1}{n}{\rm \mathbb{E}}\left\{ \left\Vert \hat{\bm{x}}-\bm{x}\right\Vert _{2}^{2}\right\} ,\;{\rm s.t.}\;\sum_{i=1}^{n}\sigma_{i}^{2}=n,\label{eq:problem-formulation}
\end{equation}
where $\hat{\bm{x}}$ is the compressed sensing reconstruction.

\subsection{Background on AMP}

The AMP framework involves a soft thresholding function and the associated
MSE analysis. Consider a scalar system $y=x+w$ where $x\sim p_{\epsilon}$
and $w\sim\mathcal{N}\left(0,\sigma^{2}\right)$. Given $y$, AMP
employs the soft thresholding function 
\begin{equation}
\hat{x}=\eta\left(y;\;\theta\right)\triangleq\begin{cases}
y-\theta & {\rm if}\; y>\theta,\\
y+\theta & {\rm if}\; y<-\theta,\\
0 & {\rm otherwise},
\end{cases}\label{eq:soft-thresholding}
\end{equation}
to estimate $x$, where $\theta\ge0$ is a threshold. Consider the
reconstruction MSE
\[
M\left(p_{\epsilon},\sigma^{2}\right)=\underset{\theta\ge0}{{\rm \inf}}\;{\rm \mathbb{E}}\left\{ \left(\hat{x}-x\right)^{2}\right\} ,
\]
where the threshold $\theta$ is optimally chosen for the given prior
distribution $p_{\epsilon}$ and noise variance $\sigma^{2}$. Introduce
the three-point mixture
\begin{equation}
p_{\epsilon,\mu}=\frac{\epsilon}{2}\delta_{-\mu}+\left(1-\epsilon\right)\delta_{0}+\frac{\epsilon}{2}\delta_{+\mu},\label{eq:least-fav-prior-1}
\end{equation}
where $\delta_{c}$ is the Delta function centered at $c$. It can
be shown that among all sparse distributions in the family of $\mathcal{F}_{\epsilon}$
(\ref{eq:prob-family-epsilon}), the (\emph{worst}) one that results
in the maximum reconstruction MSE is when $\mu=\infty$. Denote the
worst case (\emph{least favorable}) prior distribution by $p_{\epsilon}^{\#}$
($p_{\epsilon}^{\#}=p_{\epsilon,\infty}$). The associated reconstruction
MSE has the nice property 
\begin{equation}
M\left(p_{\epsilon}^{\#},\sigma^{2}\right)=\sigma^{2}M\left(p_{\epsilon}^{\#},1\right)=\sigma^{2}M^{\#}\left(\epsilon\right),\label{eq:scaling-invariant}
\end{equation}
where $M^{\#}\left(\epsilon\right)\triangleq M\left(p_{\epsilon}^{\#},1\right)$
is introduced to simplify the notations and referred to as \emph{minimax
MSE}. A closed form to compute $M^{\#}\left(\epsilon\right)$ for
an $\epsilon\in\left(0,1\right)$ has been given in \cite{journals/corr/abs-1011-4328}.
The optimal threshold is of the form $\theta=\alpha\sigma$ where
$\alpha$ is a constant only dependent on nonzero probability $\epsilon$. 
\begin{rem}
To analyse the more general case, the three-point mixture $p_{\epsilon,\mu}$
with finite $\mu$ becomes important. The associated scaling rule
is given by $M\left(p_{\epsilon,\mu},\sigma^{2}\right)=\sigma^{2}M\left(p_{\epsilon,\mu/\sigma},1\right),$
and reconstruction MSE of $\sigma^{2}=1$ also has an explicit form.
Despite the nice forms for the scalar case, the state evolution for
overall performance analysis turns out more complicated. We omit the
corresponding details due to the space constraint. 
\end{rem}
Based on the results for the scalar case, the AMP algorithm to recover
sparse $\bm{x}$ from CS measurements (\ref{eq:CS}) has been derived
\cite{journals/corr/abs-1011-4328,domamo09}: 
\begin{align}
\bm{x}^{t+1} & =\eta\left(\bm{x}^{t}+\bm{A}^{T}\bm{r}^{t};\;\bm{\theta}^{t}\right),\label{eq:typical-AMP}\\
\bm{r}^{t} & =\bm{y}-\bm{A}x^{t}+\frac{1}{m}\left\Vert \bm{x}^{t}\right\Vert _{0}\bm{r}^{t-1},\label{eq:typical-AMP2}
\end{align}
where the superscript $t$ denotes the $t$-th iteration. As $n,\, m\rightarrow\infty$
simultaneously with a constant ratio $m/n\rightarrow\delta$, a closed-form
formula to compute the minimax MSE $\frac{1}{n}{\rm \mathbb{E}}\left\{ \left\Vert \hat{\bm{x}}-\bm{x}\right\Vert _{2}^{2}\right\} $
has been derived in \cite{journals/tit/BayatiM11}. It is noteworthy
that the algorithm (\ref{eq:typical-AMP},\ref{eq:typical-AMP2})
and the analysis are based on the assumption that the matrix $\bm{A}$
contains i.i.d. Gaussian entries.

\section{\label{sec:Revised-AMP}Revised AMP with A Given Power Allocation}

When coming to power allocation, the original AMP algorithm (\ref{eq:typical-AMP},\ref{eq:typical-AMP2})
needs to be tailored. It has been assumed that a column of $\bm{A}$,
say $\bm{A}_{i}$, contains entries generated from i.i.d. ${\cal N}\left(0,\;\sigma_{i}^{2}/m\right)$.
The original AMP is not optimal any more as different columns may
have different $\ell_{2}$-norm. The revised AMP, termed as AMP.P($\bm{\epsilon}$),
is given by 
\begin{align}
\bm{x}^{t+1} & =\eta\left(\bm{x}^{t}+\bm{\Theta}^{-2}\bm{A}^{T}\bm{r}^{t};\;\bm{\Theta}^{-1}\bm{\theta}^{t}\right),\label{eq:eAMP_x}\\
\bm{r}^{t} & =\bm{y}-\bm{A}x^{t}+\frac{1}{m}\left\Vert \bm{x}^{t}\right\Vert _{0}\bm{r}^{t-1},\label{eq:eAMP_r}
\end{align}
where $\bm{\Theta}^{2}\triangleq{\rm diag}\left(\sigma_{1}^{2},\sigma_{2}^{2},...,\sigma_{n}^{2}\right)$.
The major difference from the standard one is the terms $\bm{\Theta}^{-2}$
and $\bm{\Theta}^{-1}$ in (\ref{eq:eAMP_x}). It is noteworthy that
the revised AMP is not particularly designed for the worst case though
the later analysis is.

\subsection{\label{sub:The-Derivations}Derivations}

The derivation of the AMP.P($\bm{\epsilon}$) follows from the same
idea behind the standard AMP \cite{journals/corr/abs-1011-4328}.
Describe the statistical relationship between $\bm{x}$ and $\bm{y}$
by a bipartite graph, which includes variable nodes indexed by $i\in\left[n\right]$
for variables $x_{i}$ and factor nodes indexed by $a\in\left[m\right]$
corresponding to observations $y_{a}$. Denote the message passed
from the factor node $a$ to the variable node $i$ by $r_{a\rightarrow i}^{t}$
and that from the variable node $i$ to the factor node $a$ by $x_{i\rightarrow a}^{t}$,
where the superscript $t$ denotes the $t^{th}$ iteration. It can
be verified that \cite{journals/corr/abs-1011-4328}
\begin{align}
r_{a\rightarrow i}^{t} & =y_{a}-\sum_{j\in[n]\backslash i}A_{aj}x_{j\rightarrow a}^{t},\label{eq:eAMP_r_ai}\\
x_{i\rightarrow a}^{t+1} & =\frac{1}{\sigma_{i}^{2}}\eta_{t}\left(\sum_{b\in[m]\backslash a}A_{bi}r_{b\rightarrow i}^{t}\right),\label{eq:eAMP_x_ia}
\end{align}
where for notational convenience, $\eta\left(\cdot,\theta_{t}\right)$
is simplified to $\eta_{t}\left(\cdot\right)$ henceforth. The crux
of the AMP is to approximate these messages so that the computational
complexity can be significantly reduced. 

In the approximation, only ${\cal O}\left(1\right)$ and ${\cal O}\left(n^{-1/2}\right)$
terms are kept and all smaller terms are omitted. Here, it is assume
that both $n$ and $m$ are large and $\delta\triangleq m/n$ is a
constant strictly positive. Since $A_{a,i}\sim\mathcal{N}\left(0,\sigma_{i}^{2}/m\right)$,
it is clear $A_{a,i}$ is of ${\cal O}\left(n^{-1/2}\right)$. Note
that $r_{a\rightarrow i}^{t}=y_{a}-\sum_{j\in[n]}A_{aj}x_{j\rightarrow a}^{t}+A_{ai}x_{i\rightarrow a}^{t}$
where only the last term (of ${\cal O}\left(n^{-1/2}\right)$) depends
on $i$. One can write $r_{a\rightarrow i}^{t}=r_{a}^{t}+\delta r_{a\rightarrow i}^{t}$
where $r_{a}^{t}$ is of ${\cal O}\left(1\right)$ and both $\delta r_{a\rightarrow i}^{t}$
is of ${\cal O}\left(n^{-1/2}\right)$. By similar arguments, it holds
that $x_{i\rightarrow a}^{t}=x_{i}^{t}+\delta x_{i\rightarrow a}^{t}$,
where again, $x_{i}^{t}$ is of ${\cal O}\left(1\right)$ and $\delta x_{i\rightarrow a}^{t}$
is of ${\cal O}\left(n^{-1/2}\right)$. Keeping only ${\cal O}\left(1\right)$
and ${\cal O}\left(n^{-1/2}\right)$ terms, the equations (\ref{eq:eAMP_r_ai})
and (\ref{eq:eAMP_x_ia}) become 
\begin{align}
r_{a}^{t}+\delta r_{a\rightarrow i}^{t} & =y_{a}-\sum_{j\in[n]}A_{aj}\left(x_{j}^{t}+\delta x_{j\rightarrow a}^{t}\right)+A_{ai}x_{i}^{t},\label{eq:4.4.3}\\
x_{i}^{t+1}+\delta x_{i\rightarrow a}^{t+1} & =\frac{1}{\sigma_{i}^{2}}\eta_{t}\left(\sum_{b\in[m]}A_{bi}\left(r_{b}^{t}+\delta r_{b\rightarrow i}^{t}\right)-A_{ai}r_{a}^{t}\right).\label{eq:4.4.4}
\end{align}
From (\ref{eq:4.4.3}), it is straightforward to recognize that 
\begin{align}
r_{a}^{t} & =y_{a}-\sum_{j\in[n]}A_{aj}\left(x_{j}^{t}+\delta x_{j\rightarrow a}^{t}\right);\label{eq:r_a}\\
\delta r_{a\rightarrow i}^{t} & =A_{ai}x_{i}^{t}.\label{eq:delta_r_ai}
\end{align}
By Taylor expansion of $\eta_{t}\left(\cdot\right)$, Equation (\ref{eq:4.4.4})
becomes 
\begin{align}
x_{i}^{t}+\delta x_{i\rightarrow a}^{t} & =\frac{1}{\sigma_{i}^{2}}\eta_{t}\left(\sum_{b\in[m]}A_{bi}\left(r_{b}^{t}+\delta r_{b\rightarrow i}^{t}\right)\right)+\nonumber \\
 & \quad\frac{1}{\sigma_{i}^{2}}A_{ai}r_{a}^{t}\eta'_{t}\left(\sum_{b\in[m]}A_{bi}\left(r_{b}^{t}+\delta r_{b\rightarrow i}^{t}\right)\right),\label{eq:4.4.5}
\end{align}
from which it is clear that 
\begin{align}
x_{i}^{t+1} & =\frac{1}{\sigma_{i}^{2}}\eta_{t}\left(\sum_{b\in[m]}A_{bi}\left(r_{b}^{t}+\delta r_{b\rightarrow i}^{t}\right)\right);\label{eq:x_i}\\
\delta x_{i\rightarrow a}^{t+1} & =\frac{1}{\sigma_{i}^{2}}A_{ai}r_{a}^{t}\eta'_{t}\left(\sum_{b\in[m]}A_{bi}\left(r_{b}^{t}+\delta r_{b\rightarrow i}^{t}\right)\right).\label{eq:delta_x_ia}
\end{align}
Substitute (\ref{eq:delta_r_ai}) into (\ref{eq:x_i}) and (\ref{eq:delta_x_ia})
into (\ref{eq:r_a}). Again omit the terms smaller than ${\cal O}\left(n^{-1/2}\right)$.
We have
\begin{align}
x_{i}^{t+1} & =\frac{1}{\sigma_{i}^{2}}\eta_{t}\left(\sigma_{i}^{2}x_{i}^{t}+\left(\bm{A}^{T}\bm{r}^{t}\right)_{i}\right),\label{eq:x_i_t+1}\\
r_{a}^{t} & =y_{a}-\sum_{j\in[n]}A_{aj}x_{j}^{t}+\nonumber \\
 & \quad\sum_{j\in[n]}\frac{A_{aj}^{2}}{\sigma_{j}^{2}}\eta'_{t-1}\left(\sigma_{j}^{2}x_{j}^{t-1}+\left(\bm{A}^{T}\bm{r}^{t-1}\right)_{j}\right)r_{a}^{t-1}.\label{eq:r_a_t}
\end{align}
Note that for large $n$, $\bm{A}_{aj}^{2}\approx\sigma_{j}^{2}/m$.
The last term on the right hand side of Equation (\ref{eq:r_a_t})
can be approximated as
\begin{equation}
\sum_{j\in[n]}\frac{1}{m}\eta'_{t-1}\left(\sigma_{j}^{2}x_{j}^{t-1}+\left(\bm{A}^{T}r^{t-1}\right)_{j}\right)r_{a}^{t-1}=\frac{1}{m}\left\Vert \bm{x}^{t}\right\Vert _{0}r_{a}^{t-1}.\label{eq:b_t}
\end{equation}
Combine Equation (\ref{eq:x_i_t+1}), (\ref{eq:r_a_t}), and (\ref{eq:b_t}).
We obtain the AMP.P($\bm{\epsilon}$) iterations described by (\ref{eq:eAMP_x})
and (\ref{eq:eAMP_r}).

\section{\label{sec:Reconstruction-MSE}Reconstruction MSE and A Heuristic
Derivation}

We analyze the MSE performance of AMP.P($\bm{\epsilon}$). We focus
on the minimax MSE as the analysis can be highly simplified thanks
to the property (\ref{eq:scaling-invariant}). As the rigorous analysis
\cite{journals/tit/BayatiM11} is still too arduous, we follow the
heuristic proof in \cite{journals/corr/abs-1011-4328} which is much
easier to describe and highlights the key ideas. 

The main results can be summarized as follows. Consider the asymptotic
region where $\left(m,n\right)\rightarrow\infty$ simultaneously with
a constant ratio $m/n\rightarrow\delta$. Assume the block sparsity
structure described before with $\left|\mathcal{I}_{i}\right|/n\rightarrow c_{i}$
for some constant $c_{i}$. Consider the least favorable prior $p^{\#}\left(\epsilon_{i}\right)$,
$i\in\left[n\right]$, and suppose that ${\rm lim}_{\left(m,n\right)\rightarrow\infty}\;\frac{1}{m}\overset{n}{\underset{i=1}{\sum}}M^{\#}\left(\epsilon_{i}\right)<1$.
The minimax MSE of the revised AMP algorithm is given by
\begin{equation}
\frac{1}{n}{\rm \mathbb{E}}\left\{ \left\Vert \hat{\bm{x}}-\bm{x}\right\Vert _{2}^{2}\right\} \doteq\frac{\frac{1}{n}\overset{n}{\underset{i=1}{\sum}}M^{\#}\left(\epsilon_{i}\right)/\sigma_{i}^{2}}{1-\frac{1}{m}\overset{n}{\underset{i=1}{\sum}}M^{\#}\left(\epsilon_{i}\right)}\sigma^{2},\label{eq:minimax-reconstruction-error}
\end{equation}
where the symbol $\doteq$ denotes the equality in the aforementioned
asymptotic region. 
\begin{rem}[\emph{Relation with the Previous Result}]
\emph{} Consider the uniformly sparse signal $\bm{x}$ with $\epsilon_{i}=\epsilon_{j}$
for all $i,j\in\left[n\right]$. The minimax MSE in (\ref{eq:minimax-reconstruction-error})
becomes 
\[
\frac{M^{\#}\left(\epsilon\right)}{1-M^{\#}\left(\epsilon\right)/\delta}\sigma^{2},
\]
which is consist with the result given in \cite{journals/tit/DonohoMM11}.
\end{rem}
\vspace{0cm}
\begin{rem}[\emph{Phase-Transition for the Noiseless Case}]
\emph{} For noiseless case, $\sigma^{2}=0$. Consider the same asymptotic
region as specified before with additionally $\Sigma_{i=1}^{n}\epsilon_{i}/m\rightarrow\rho$.
The phase-transition curve that separates the sparsity-undersampling
($\rho-\delta$) plane \cite{journals/corr/abs-1011-4328} is given
by 
\[
\frac{1}{n}\overset{n}{\underset{i=1}{\sum}}M^{\#}\left(\epsilon_{i}\right)\doteq\delta.
\]
That is, the reconstruction is exact if and only if $\frac{1}{n}\sum M^{\#}\left(\epsilon\right)<\delta$.
This result is consistent with the one in \cite{5712950}. Furthermore,
note the phase transition curve is independent of $\sigma_{i}^{2}$.
It can be concluded that power allocation will not affect the phase
transition curve when there is no noise. 
\end{rem}
\vspace{0cm}

\subsection{The heuristic derivation}

The heuristic derivation of (\ref{eq:minimax-reconstruction-error})
starts with the iterative algorithm that the term $\frac{1}{m}\left\Vert \bm{x}^{t}\right\Vert _{0}\bm{r}^{t-1}$
in (\ref{eq:eAMP_r}) is omitted, i.e.,
\begin{align}
\bm{x}^{t+1} & =\eta_{t}\left(\bm{x}^{t}+\bm{\Theta}^{-2}\bm{A}^{T}\bm{r}^{t}\right),\label{eq:IT_x}\\
\bm{r}^{t} & =\bm{y}-\bm{A}\bm{x}^{t}.
\end{align}
Meantime, it also poses an artificial assumption that the matrix $\bm{A}$
at different iterations are independently generated. Note in reality
the matrix $\bm{A}$ is fixed for all the iterations. The heuristic
derivation gives the correct analysis as adding term (\ref{eq:b_t})
will make the residue noise from different iterations independent.

To proceed, the input of the thresholding function in (\ref{eq:IT_x})
can be written as

\begin{align}
\bm{x}^{t}+\bm{\Theta}^{-2}\bm{A}^{T}\bm{r}^{t} & =\bm{x}^{t}+\bm{\Theta}^{-2}\bm{A}^{T}\left(\bm{y}-\bm{A}\bm{x}^{t}\right)\nonumber \\
 & =\bm{x}+\bm{e}^{t},\label{eq:denoiser}
\end{align}
where $\bm{e}^{t}\triangleq\left(\bm{\Theta}^{-2}\bm{A}^{T}\bm{A}-\bm{I}\right)\left(\bm{x}-\bm{x}^{t}\right)+\bm{\Theta}^{-2}\bm{A}^{T}\bm{w}$.
The explicit form of the matrix $\left(\bm{\Theta}^{-2}\bm{A}^{T}\bm{A}-\bm{I}\right)$
in $\bm{e}^{t}$ is
\[
\left[\begin{array}{ccc}
\sigma_{1}^{-2}\bm{A}_{1}^{T}\bm{A}_{1}-1 & \sigma_{1}^{-2}\bm{A}_{1}^{T}\bm{A}_{2} & \cdots\\
\sigma_{2}^{-2}\bm{A}_{2}^{T}\bm{A}_{1} & \sigma_{2}^{-2}\bm{A}_{2}^{T}\bm{A}_{2}-1 & \cdots\\
\vdots & \vdots & \ddots
\end{array}\right].
\]
It can be verified that each diagonal entry $\sigma_{i}^{-2}\bm{A}_{i}^{T}\bm{A}_{i}-1$
is approximately normal with zero mean and variance $2/m$; each off-diagonal
entry $\sigma_{i}^{-2}\bm{A}_{i}^{T}\bm{A}_{j}$, $i\neq j$, has
zero mean and variance $\sigma_{i}^{-2}\sigma_{j}^{2}/m$. By the
fact that $\bm{w}\sim{\cal N}\left(0,\;\sigma^{2}\bm{I}\right)$,
the following properties hold: 1) $\mathbb{E}\left\{ e_{i}^{t}\right\} =0$;
2) $\mathbb{E}\left\{ e_{i}^{t}e_{j}^{t}\right\} =0,$ $i\neq j$;
3) for large $n$, define $\tilde{\tau}_{t,i}^{2}\triangleq\mathbb{E}\left\{ \left|e_{i}^{t}\right|^{2}\right\} $,
where 
\begin{align*}
\mathbb{E}\left\{ \left|e_{i}^{t}\right|^{2}\right\}  & \doteq\frac{1}{\sigma_{i}^{2}}\left(\overset{n}{\underset{j=1}{\sum}}\frac{\sigma_{j}^{2}}{m}\mathbb{E}\left\{ \left|x_{j}-x_{j}^{t}\right|_{2}^{2}\right\} +\sigma^{2}\right).
\end{align*}
This helps in quantifying the MSE at the $\left(t+1\right)^{th}$
iteration: 
\begin{align*}
\tilde{\tau}_{t+1,i}^{2} & \doteq\frac{1}{\sigma_{i}^{2}}\left(\overset{n}{\underset{j=1}{\sum}}\frac{\sigma_{j}^{2}}{m}\mathbb{E}\left\{ \left|x_{j}-\eta_{t}\left(x_{j}+e_{j}^{t}\right)\right|_{2}^{2}\right\} +\sigma^{2}\right).
\end{align*}
From the definition of $M^{\#}\left(\epsilon_{j}\right)$ in (\ref{eq:scaling-invariant}),
\begin{equation}
\mathbb{E}\left\{ \left|x_{j}-\eta_{t}\left(x_{j}+e_{j}^{t}\right)\right|_{2}^{2}\right\} =M^{\#}\left(\epsilon_{j}\right)\tilde{\tau}_{t,j}^{2}.\label{eq:minimax-reconst-error-i}
\end{equation}
As a result, when the steady state ($\tilde{\tau}_{t,j}=\tilde{\tau}_{t+1,j}$)
is reached, 
\begin{equation}
\tilde{\tau}_{i}^{2}\doteq\frac{1}{\sigma_{i}^{2}}\left(\frac{1}{m}\overset{n}{\underset{j=1}{\sum}}\sigma_{j}^{2}M^{\#}\left(\epsilon_{j}\right)\tilde{\tau}_{j}^{2}+\sigma^{2}\right),\; i\in[n].\label{eq:state-evolution}
\end{equation}
The explicit form to compute $\tilde{\tau}_{i}^{2}$ can be computed
by observing that for all $i\in\left[n\right]$, $\tilde{\tau}_{i}^{2}\sigma_{i}^{2}=\overset{n}{\underset{j=1}{\sum}}\frac{\sigma_{j}^{2}}{m}M^{\#}\left(\epsilon_{j}\right)\tilde{\tau}_{j}^{2}+\sigma^{2}$
which is a constant independent of $i$. Hence,
\begin{align}
\tilde{\tau}_{i}^{2} & \doteq\frac{\sigma^{2}}{\sigma_{i}^{2}}\cdot\frac{1}{1-\frac{1}{m}\overset{n}{\underset{i=1}{\sum}}M^{\#}\left(\epsilon_{i}\right)},\; i\in[n].\label{eq:state-evolution2}
\end{align}
Combine (\ref{eq:state-evolution2}) with the state evolution (\ref{eq:minimax-reconst-error-i}).
 We obtain 
\begin{align*}
\frac{1}{n}{\rm \mathbb{E}}\left\{ \left\Vert \hat{\bm{x}}-\bm{x}\right\Vert _{2}^{2}\right\} \doteq\frac{1}{n}\overset{n}{\underset{i=1}{\sum}}M^{\#}\left(\epsilon_{i}\right)\tilde{\tau}_{i}^{2} & ,
\end{align*}
which gives (\ref{eq:minimax-reconstruction-error}).

\section{Optimal Power Allocation}

Based on the derived minimax MSE, the optimal power allocation can
be achieved. In particular, the power allocation can be formulated
as a constrained optimization problem
\[
\underset{\sigma_{i},\; i\in[n]}{{\rm min}}\;\frac{\frac{1}{n}\overset{n}{\underset{i=1}{\sum}}M^{\#}\left(\epsilon_{i}\right)/\sigma_{i}^{2}}{1-\frac{1}{m}\overset{n}{\underset{i=1}{\sum}}M^{\#}\left(\epsilon_{i}\right)}\sigma^{2},\;{\rm s.t.}\;\overset{n}{\underset{i=1}{\sum}}\sigma_{i}^{2}=n.
\]
As $\sigma_{i}^{2}$'s are the only variables, focus on the numerator
of the objective function. By the \emph{Cauchy-Schwarz} inequality,
one has
\begin{align}
\overset{n}{\underset{i=1}{\sum}}\frac{M^{\#}\left(\epsilon_{i}\right)}{\sigma_{i}^{2}} & =\overset{n}{\underset{i=1}{\sum}}\frac{M^{\#}\left(\epsilon_{i}\right)}{\sigma_{i}^{2}}\cdot\frac{1}{n}\overset{n}{\underset{i=1}{\sum}}\sigma_{i}^{2}\nonumber \\
 & \geq\frac{1}{n}\left(\overset{n}{\underset{i=1}{\sum}}\sqrt{M^{\#}\left(\epsilon_{i}\right)}\right)^{2},
\end{align}
where the equality holds if and only if $\sqrt{M^{\#}\left(\epsilon_{i}\right)}=c\sigma_{i}^{2}$
for some constant $c$. Recall the total power constraint $\sum\sigma_{i}^{2}=n$.
The constant $c$ can be characterized and the optimal power allocation
is given by 
\begin{equation}
\sigma_{i}^{2}=\frac{\sqrt{M^{\#}\left(\epsilon_{i}\right)}}{\frac{1}{n}\overset{n}{\underset{i=1}{\sum}}\sqrt{M^{\#}\left(\epsilon_{i}\right)}},\; i\in[n].\label{eq:optimum-results}
\end{equation}

\section{Discussion}

\begin{figure}
\centering{}\includegraphics[scale=0.33]{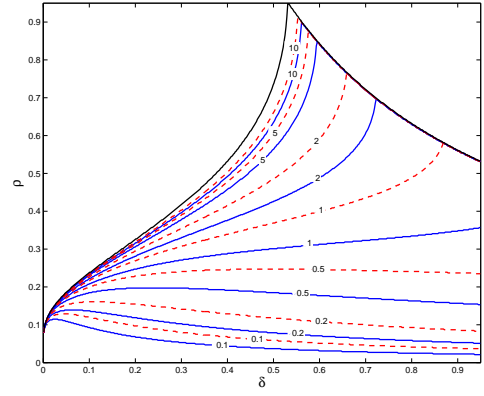}\caption{\label{fig:theoretical-curve}Reconstruction error contours for a
sparse signal with two even-length blocks where the sparsity ratio
$\epsilon^{\left(1\right)}/\epsilon^{\left(2\right)}=100$. The blue
solid lines and the red dashed lines respectively present the minimax
MSEs $\left\{ 0.1,\,0.2,\,0.5,\,1,\,2,\,5,\,10\right\} $ before and
after the power allocation. The phase-transition curve for noiseless
case is given by the black line. The upper right curved area is the
inadmissible area under the sparsity ratio 100.}
\end{figure}

\subsection{Theoretical Reconstruction Error}

For theoretical demonstration of the effects of power allocation,
we assume that the unknown sparse signal can be divided into two even-length
blocks where the sparsity ratio is given by $\epsilon^{\left(1\right)}/\epsilon^{\left(2\right)}=100$.
Consider the least favorable prior $p_{\epsilon^{\left(1\right)}}^{\#}$
and $p_{\epsilon^{\left(2\right)}}^{\#}$. Normalize the noise variance
by setting $\sigma^{2}=1$. Let $\delta=m/n$ and $\rho=\frac{1}{m}\sum\epsilon_{i}$.
In Fig. \ref{fig:theoretical-curve}, the minimax MSE contours before
and after the power allocation are respectively given by blue solid
lines and red dashed lines. The phase-transition curve for noiseless
case is given by the black line. We see that for the all pairs of
$(\rho,\;\delta)$ under the phase-transition curve, the obtained
reconstruction errors decreased after power allocation. Above the
phase-transition bound the state evolution does not converge. The
reconstruction error goes to infinity.

\begin{figure}
\centering{}\includegraphics[scale=0.47]{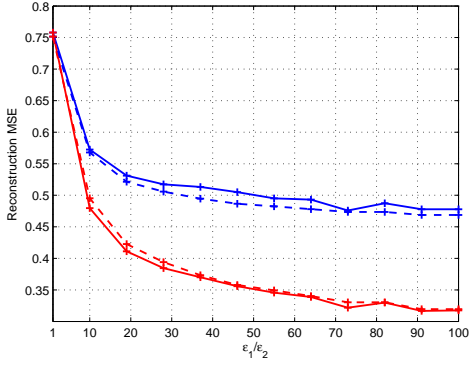}\caption{\label{fig:Mse-against-xi}MSE against sparsity ratio for sparse signals
with two even-length blocks. Blue and red solid lines are the MSE
before and after power allocation. Dashed lines are the corresponding
theoretical prediction. }
\end{figure}
\begin{figure}
\centering{}\includegraphics[scale=0.47]{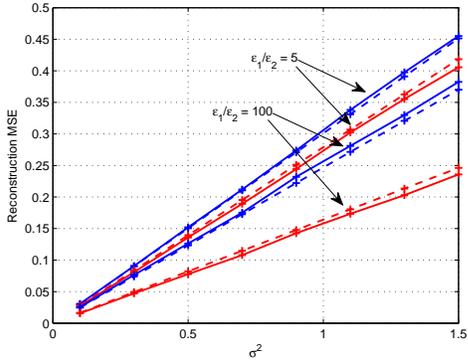}\caption{\label{fig:Mse-against-sigma2}MSE against noise variance for sparse
signals with two even-length blocks. Number of realizations is 100.
Blue and red solid lines are MSE curves before and after power allocation.
Dashed lines are the corresponding theoretical prediction. }
\end{figure}

\subsection{Empirical Studies}

The least favorable prior involves Diracs at $\pm\infty$. It is not
practical to numerically generate a sparse signal from such a prior.
To avoid this difficulty, the authors of \cite{journals/tit/DonohoMM11}
defined the so called $a$\emph{-least favorable} prior as the distribution
$p_{\epsilon,\mu}\in{\cal F}_{\epsilon}$ such that the corresponding
MSE satisfies $M_{a}\left(\epsilon\right)=\left(1-a\right)M^{\#}\left(\epsilon\right)$,
where $0<a\ll1$. Given an $a$, the value of $\mu$ can be computed
via the explicit form of the MSE of the three-point mixture (see the
journal version of this paper for more details). 

We set $a=0.02$ which is the same as that in \cite{journals/tit/DonohoMM11}.
Let $m=2000$ and $n=4000$. Assume a sparse signal with two even-length
blocks, i.e., $n_{1}=n_{2}=n/2$. The sparsity ratio is defined as
$\epsilon^{\left(1\right)}/\epsilon^{\left(2\right)}$. The signal
$\bm{x}$ is randomly generated (100 realizations) from the sparse
prior. For each realization, the AMP.P($\bm{\epsilon}$) algorithm
is applied for reconstruction to obtain $\hat{\bm{x}}$. In Fig. \ref{fig:Mse-against-xi},
we fix $\rho=0.18$ but vary the sparsity ratio $\epsilon^{\left(1\right)}/\epsilon^{\left(2\right)}$.
We compare the reconstruction MSE $\left\Vert \hat{\bm{x}}-\bm{x}\right\Vert _{2}^{2}/n$.
From the presented results, the average MSE after power allocation
is always smaller. The performance gain becomes larger when the sparsity
ratio increases. Theoretical predictions drawn as dashed curves are
very close to the curves obtained from simulations. In Fig. \ref{fig:Mse-against-sigma2},
we aim to demonstrate the linear relationship between the reconstruction
MSE and the noise variance, predicted by (\ref{eq:minimax-reconstruction-error}).
The settings are the same to those for Fig. \ref{fig:Mse-against-xi}
except that $\rho=0.1$ and $\epsilon^{\left(1\right)}/\epsilon^{\left(2\right)}=5$
and 100. From the simulations, the linear relationship is confirmed.

\section{Conclusion}

In this paper we consider non-uniformly sparse signals. We first show
in the presence of noise, i.i.d. Gaussian random measurement matrix
may not be optimal in minimizing the reconstruction MSE. Then we considered
how to allocate a given total power across the columns of the measurement
matrix. Given a power allocation, we derived the AMP.P($\bm{\epsilon}$)
algorithm, and quantitatively analyzed the corresponding minimax MSE.
Based on it, the optimal power allocation policy has been identified.
Both theoretical and empirical results are presented with the clear
consistency and verified the performance gain.

\bibliographystyle{IEEEtran}
\bibliography{Zhao-References}

\end{document}